\documentclass[sigconf,nonacm]{acmart}

\usepackage{booktabs} 

\setcopyright{acmcopyright}

\acmArticle{4}
\acmPrice{15.00}

\begin{document}
\title{Untangling the Spaghetti Code in Game Development: A Review of Challenges and Academic Solutions}

\renewcommand{\shorttitle}{Code Smells and Technical Debt in Game Projects and Academic Solutions}

\author{Esdras Caleb Oliveira Silva}
\email{esdrascaleb@gmail.com}
\affiliation{%
  \institution{UFRN}
  \city{Natal}
  \country{Brasil}
}

\author{Lyrene Fernandes da Silva}
\email{lyrene@gmail.com}
\affiliation{%
  \institution{UFRN}
  \city{Natal}
  \country{Brasil}
}

\renewcommand{\shortauthors}{E. C. O. Silva and L. F. da Silva}

\begin{abstract}
The code quality in the game domain is perceived as less structured and more convoluted than in other domains, which results in lower maintainability and higher bug frequency. In this work, we conduct a systematic literature review to address four main research questions: (1) how the incidence of code smells and technical debt differs in game development compared to other software domains; (2) what academic solutions are proposed to mitigate these code quality issues; (3) whether researchers collaborate effectively with developers to implement these solutions; and (4) if the proposed solutions are sufficient to tackle the diagnosed problems. Thirty-four works were mapped and analyzed. We found that the game domain is affected by a short-term mentality, continuous requirements changes, reduced code reuse, and minimal to no automated testing. As potential solutions, we identified tools and models to detect code smells, along with testing practices and product lines adapted for game development. Although these solutions exist, they are not widely adopted by practitioners, highlighting the need to investigate barriers to their integration.
\end{abstract}

%
%


\maketitle

\section{Introduction}
While bugs are common across various domains, they are perceived as a recurrent issue specifically in the game domain. Games are increasingly perceived as more bugged by the players~\cite{cbr2024}, and developers have admitted that games are released in a bugged state~\cite{zhang2016elevator}. 

Bugs are not always a direct consequence of smells~\cite{piotrowski2020software}, but their existence impacts the work needed to fix them~\cite{palomba2017code_smells}. The recent research in technical debt in Twitter~\cite{alfayez2024technical} shows that the game domain with the most interaction has more likes and retweets. Paradoxically, tools for detecting code smells are available and increasingly adopted~\cite{pereira2022code}, as does the amount of academic research in the domain~\cite{chueca2024consolidation}. 

The problem has long been known~\cite{zhang2016elevator}, but even being a multi-billion-dollar business enterprise~\cite{liang2022analysis}, that would benefit from maintaining high-quality code to ensure fewer problems. The industry is still plagued with flawed launches impacting the players' confidence in them~\cite{nowicki2021time}. Is there an inherent problem in game development projects? Is there relevant research and tools to mitigate them? Why do solutions appear to not work in the game developer community?

Through this systematic review, we examine whether the game development domain is particularly susceptible to code smells and explore the solutions proposed in academic literature to tackle this issue. The objective is to identify the current academic state of this problem and provide findings to inform future research. We hope to support future developments that will result in fewer bugs and a better gaming experience for players.

This paper is structured as follows: Section 2 presents some related systematic reviews and their contributions. Section 3 presents the methodology employed for the review and explains how the BART-MNLI model~\cite{lewis2020bart} was utilized to include the selected works. Section 4 discusses the results and includes a mapping of the works. Section 5 presents and discusses our findings in response to the research questions. Section 6 makes remarks on the limitations of our study. Finally, Section 7 concludes our work, providing suggestions for future research.

\section{Related Reviews}
In table~\ref{tab:reviews} we present six related systematic reviews, they are ordered in chronological publication date. The reviews by ~\cite{cairo2018impact,piotrowski2020software} present the relationship between code smells and software defects and bugs. Which is the premise used for this systematic review, as we search for code smells and technical debt solutions as a way to solve the present problem of defects in games~\cite{zhang2016elevator}. 

Piotrowski et al. ~\cite{piotrowski2020software} present the relation between technical debt and code smells, in our paper we search for both because of this relation. We also know that games have strict time constraints during development, as the review from ~\cite{becker2018trade} shows how this characteristic impacts the technical debt of this project.

Pereira et al~\cite{pereira2022code} shows ways to detect smells making a review of the most used techniques. In our review, we found works that use similar methods. Chueca et al.~\cite{chueca2024consolidation} points to the need for more systematic reviews in the Software Engineering in the game domain and how solutions to the development circle are in need.

\begin{table}[!t]
  \caption{Related Reviews}
  \label{tab:reviews}
  \begin{tabular}{c c p{1in} p{1in}}
    \toprule
    Ref. & Year & Focus & Findings \\
    \midrule
    ~\cite{cairo2018impact} & 2018 & Smells and Bugs in Software & Smell Impact on software bugs \\ \hline

    ~\cite{becker2018trade} & 2018 & Time Constraint Impact Technical Debt & The role of time limits in the technical quality\\ \hline

    ~\cite{piotrowski2020software} & 2020 & Relation of Smells and Defects & The role of some types of smells in predicting software defects\\ \hline

    ~\cite{das2022technical} & 2022 & Relation of Code Smells and Technical Debt & How smells are used to detect technical debt\\ \hline
    
    ~\cite{pereira2022code} & 2022 & Smells detection & Ways to find smells in code \\ \hline

    ~\cite{chueca2024consolidation} & 2024 & Game Software Engineer & The need of research in Game Software Engineer field \\ \hline
   \bottomrule
    \end{tabular}
\end{table}

\section{Method}
This research is according to the guidelines of \mbox{Kitchenham} and \mbox{Charters}~\cite{kitchenham2007guidelines} and its general objective is to provide a comprehensive overview of code quality issues in game development, including the occurrence of code smells and technical debt, the solutions presented in academic research, and the involvement of the development community in addressing these challenges.

\subsection{Research Questions}
In this review, we seek to answer the following research questions: 
\begin{itemize}
    \item \textbf{RQ1:} How does the incidence of code smells and technical debt differ in the Game domain from other domains? 
    \item \textbf{RQ2:} What solutions are the researchers presenting to the code smells and anti-patterns in the game domain?
    \item \textbf{RQ3:} Are researchers collaborating effectively with developers to implement these solutions?
    \item \textbf{RQ4:} Are the found solutions to the code quality problem in the game domain sufficient to tackle the diagnosed problems?
\end{itemize}

In \textbf{RQ1}, we propose to investigate how is the code quality in the game area and the prevalence of smells and anti-patterns in it. With this, we hope to find the causes that leave the players' perception that games are more bugged today.

In \textbf{RQ2}, we check the solutions generated by research and what results they are gathering. We expect to find which solutions are used in the game domain, their applications, and new possible strategies.

In \textbf{RQ3}, we check if the data acquisition and the solutions are being made with the game development community in mind. If they get data from active developers and if the solutions are being tested with an active community of developers. Solutions need to be compatible with the actual workflow of the game developer community to make an impact in the current situation.

In \textbf{RQ4}, we investigate if the RQ1 pointed particularities are being answered by RQ2. And point to other possible sources of solutions to the particularities in the game domain found in RQ1.

\subsection{Consulted Databases}

We select as sources of our articles Scopus, ScienceDirect, IEEE, Web Of Science, and ACM as they were commonly used in systematic reviews~\cite{chueca2024consolidation,kitchenham2007guidelines}. We also include Google Scholar as an additional source of data~\cite{piasecki2018google}, this decision was made to find works that are not visible in academic databases but could have important insights into our research.

Google Scholar has previously been compared to other sources of data, and it could have problems getting the correct number of citations~\cite{sauvayre2022types}. But the number of works available~\cite{martin2021google} and its search engine capabilities~\cite{gusenbauer2019google}, make it a valuable source in search of additional data.

\subsection{Search Terms}
To find the search terms used PICO (Population, Intervention, Comparison, Outcome) as suggested by Guidelines~\cite{kitchenham2007guidelines}, but in this review, we are not searching for a theme but for a problem inside a domain. Thus, PICO is modeled to fit our needs:
\begin{itemize}
\item \textbf{Population} is the Game Domain or the Game development community.
\item \textbf{Intervention} We are searching for code problems, such as code smells, anti-patterns, and technical debt.
\item \textbf{Comparison} We are searching for solutions to the code problems, such as patterns and other techniques to the code smells.
\item \textbf{Outcome} better code, better manageability for the code fewer bugs.
\end{itemize}

\begin{table}[!t]
  \caption{Search Strings for database}
  \label{tab:strings}
  \begin{tabular}{p{0.8in} p{2.2in}}
    \toprule
    database & Search String\\
    \midrule
    ACM DL & game AND ( "code smell" OR "code smells" OR "bad smell" OR "bad smells" OR "\mbox{technical} debt" OR "\mbox{antipattern}" OR "\mbox{antipatterns}" ) AND ( "test" OR "design \mbox{patterns}" OR "\mbox{design} pattern")  \\ \hline
    IEEE Xplore \& Scopus \& WebOfScience& game AND ( "code smell" OR "bad smell" OR "technical debt" OR "antipattern" ) \\ \hline
    Science Direct & game AND ( "code smell" OR "bad smell" OR "technical debt" OR "antipattern" ) AND ( "test" OR "design patterns" )
 \\ \hline
    Google Scholar & ("game development" OR "game \mbox{programming}" OR "video game" OR "game software") AND ( "code smell" OR "bad smell" OR "\mbox{technical} debt" OR "antipattern" ) AND ( "test" OR "design pattern" ) -gamification \\
   \bottomrule
    \end{tabular}
\end{table}

The search strings were pilot-tested and tailored according to the specific syntax and indexing characteristics of each database. Broad strings were applied to databases returning fewer studies (e.g., IEEE Xplore, Web of Science, and Scopus) to maximize recall. Conversely, Google Scholar required a more targeted query with explicit exclusions (e.g., \texttt{-gamification}) to filter out studies focused on gamifying code smell detection rather than game software development. The full set of database-specific search strings is summarized in Table~\ref{tab:strings}.

\subsection{Search Process}
\begin{figure}
\includegraphics[width=2in]{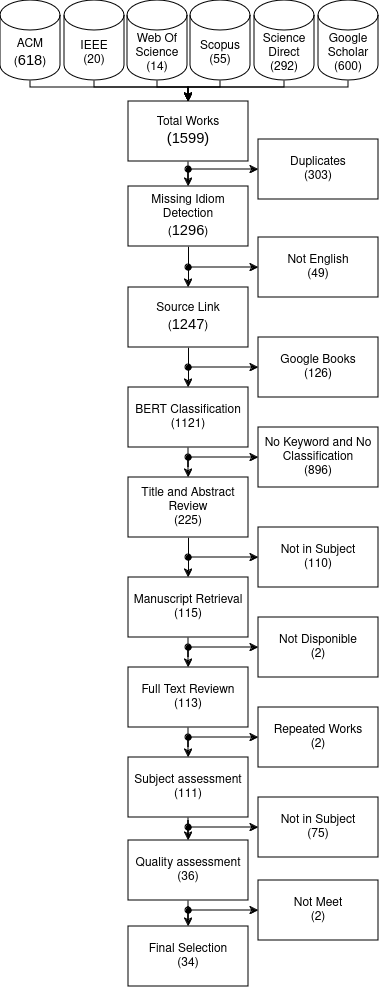}
\caption{Search Process}
\label{fig:search}
\end{figure}

A table was created with all article metadata to aid the selection and exclusion of works with the criteria. The table was created with a Python open-source script from repo~\cite {esdrascaleb2024}. The process followed is shown in the figure~\ref{fig:search}. The number of works returned from the database was 618 from ACM, 20 from IEEE, 14 From Web Of Science, 55 from Scopus, 292 from Science Direct, and 600 from Google Scholar. In total 1599 works were found. The literature search was completed on August 15, 2024, representing the cutoff date for this review; studies published after this date were not included. The table contains the fields, title, abstract, origin, URL, and keywords (which combines all keywords fields like author keywords).

\subsubsection{Exclusion and Inclusion Criteria}
To select the studies the defined exclusion (EC) and inclusion (IC) criteria were:
\begin{itemize}
\item EC1 - Duplicated Works
\item EC2 - Non-English works
\item EC3 - Books from Google results with no academic access
\item IC1 - Has "Game" in keywords field \textbf{OR} a score of at least 50 for the title or abstract given by the LLM Model (BART-MNLI) 
\item IC2 - Subject detected or doubt after abstract or title manual review
\item  IC3 - Studies that are about the review subject
\end{itemize}

The \textbf{EC1} was the removal of duplicated files, 303 files were removed with this. The \textbf{EC2} was to only consider works on English language. As not all works from Google Scholar have the metadata about the language, the Python library \textit{langdetect}~\cite{langdetect} was used to detect the language of the work. The \textit{langdetect} is based on the n-gram frequency statistics for text classification introduced in Cavnar et al~\cite{cavnar1994n}. Using the language detection and the metadata information 49 works were excluded. 

Some of the data retrieved from Google Scholar consists of books that are not openly accessible. Therefore, they could not be analyzed, so the \textbf{EC3} involved removing data sourced from Google Books. In which 126 works were removed.

After the three exclusion criteria, an inclusion criterion was formulated to select only the papers in the desired domain. The criterion was to include a work if:
\begin{enumerate}
    \item In the keywords field, the word ``Game'' is present.
    \item BART-MNLI model zero-shot classification entailment probability score for the title or abstract, evaluated against the candidate label ``Game Development'', is greater than or equal to 0.50 (50\%).
\end{enumerate}

The first criterion follows standard systematic review practices (e.g.,~\cite{chueca2024consolidation}), checking whether the keyword field explicitly includes ``Game''. However, relying solely on explicit keywords is insufficient because many records (particularly from Google Scholar) lack indexed keyword metadata, while others produce off-topic matches. To ensure both high recall and semantic relevance, the second criterion incorporates automated Natural Language Inference (NLI).

We utilized the \textit{facebook/bart-large-mnli} model~\cite{lewis2020bart,williams2018broad} via the Hugging Face zero-shot classification pipeline~\cite{mann2020language,wolf2019huggingface} without additional fine-tuning to guarantee replicability. The model computes NLI premise-hypothesis entailment probabilities for the candidate label ``Game Development'' against the title and abstract of each study, generating two distinct score fields (BART-MNLI title score and abstract score).

To validate the safety of the 0.50 entailment threshold, we conducted a full manual audit of all excluded papers that contained the term ``game'' in their metadata. Abstract-level review of these filtered works confirmed that all excluded items were indeed off-topic, no false negatives were identified within the manually audited excluded records. In comparison, relying strictly on basic keyword matching for ``game'' across title, abstract, and keywords identified 252 candidate works, many of which proved irrelevant upon inspection.

The title and the abstract were read to check if they matched one of the inclusion criteria. If there was doubt the paper was included to be fully read when it passed at least one of the following criteria:
\begin{enumerate}
\item Works contain information about code smells or technical debt in the game domain.
\item The study brings a solution to code smells, technical debt, or its consequences to be used in the game domain.
\item The study analyzes data from commercial game companies that could elucidate the matter.
\end{enumerate}

115 works were selected to be read. When the articles were searched for retrieval, two of them were not accessible. So 113 works had their title and abstracts read, of which 36 were considered relevant to this study. Two works were considered duplicated works and did not bring any useful complementary information so they were also excluded.

\subsubsection{Quality Assessment}
A quality assessment was made with the remaining works, it followed the same principles from Galster et al~\cite{galster2013variability} and Chueca et al~\cite{chueca2024consolidation}. With the change to use only a 2-point scale of yes or no in the instrument. The quality assessment questions were the same in both referred studies:
\begin{enumerate}
\item Is there a rationale provided for why the study was undertaken?
\item Is there an adequate description of the context in which the research was carried out?
\item Is there a justification and description for the research design?
\item Is there a clear statement of the findings, including data that supports findings?
\item Did the researcher(s) critically examine their own role, potential bias, and influence during the study
\item Are the limitations and credibility of the study discussed explicitly?
\end{enumerate}
To meet our quality assessment a study must have four points, by answering yes to at least four of the six questions. Two studies were excluded from the final selection due to their low scores of 1 and 2 points, respectively. Therefore, from the 34 studies, 4 achieved the maximum quality, 21 had 5 points and 9 had 4 points. One study is a systematic review and the other 33 are primary studies. The systematic review was kept as it offers a general overview of a related topic. Their findings and contents will be discussed in the next section.

\section{Results}

Table~\ref{tab:studies} summarizes how the selected studies collected their data, while Table~\ref{tab:synthesis} provides a synthesis mapping the main proposed solutions to our research questions. Among the mapped literature, 20 works presented a code analysis and 19 conducted experiments with code, whereas 11 utilized interviews and surveys. The study~\cite{borrelli2020detecting} belongs to both groups, representing the most comprehensive approach by validating its findings through both experiments and a survey. Importantly, 57.6\% of the studies had a direct connection with the community, either through consulting (11) or corporate partnerships (8), which provides key evidence for \textbf{RQ3}. In the next subsections, we analyze these studies grouped by their main contribution.

\begin{table*}[!t]
  \caption{Overview of Selected Studies}
  \label{tab:studies}
  \begin{tabular}{c c c c c c c c c}
    \toprule
    Ref. & Year & Contribution & Code Analyzed & Industry Assessment & Fieldwork & Quality & Source & Type  \\
    \midrule
    ~\cite{gustafsson2011model} & 2011 & Management & - & Partner & Interview & 5 & Scholar & Master Thesis \\
    ~\cite{murphy2014cowboys} & 2014 & Domain Spec. & - & Consulting & Sur. \& Int. & 5 & ACM & Article \\
    ~\cite{olsson2014evaluation} & 2014 & Smell Detec. & Proprietary & Affiliate & Experiment & 4 & ACM & Article \\
    ~\cite{gilliland2015empirical} & 2015 & Prod. Lines & Proprietary & Partner & - & 4 & Scholar & Master Thesis \\
    ~\cite{tapp2015employee} & 2015 & Management & - & Partner & Interview & 5 & Scholar & Master Thesis \\
    ~\cite{hermans2016code} & 2016 & Smell Impact & - & - & Experiment & 5 & IEEE & Article \\
    ~\cite{pereira2016code} & 2016 & Domain Spec. & OpenSource & - & Experiment & 5 & IEEE & Article \\
    ~\cite{gibson2017mind} & 2017 & Domain Spec. & - & Consulting & Interview & 4 & Scholar & Article \\
    ~\cite{kruger2018apo} & 2018 & Smell Detec. & OpenSource & - & Experiment & 5 & WebOfScience & Article \\
    ~\cite{aakesson2019migrating} & 2019 & Prod. Lines & OpenSource & - & Experiment & 5 & ACM & Article \\
    ~\cite{borg2019video} & 2019 & Domain Spec. & - & Consulting & Survey & 5 & Scholar & Article \\
    ~\cite{debbiche2019migrating} & 2019 & Prod. Lines & OpenSource & - & Experiment & 5 & ACM & Article \\
    ~\cite{khanve2019existing} & 2019 & Smell Detec. & OpenSource & - & Experiment & 4 & ACM & Article \\
    ~\cite{racheva2019testing} & 2019 & Tests & - & Partner & Interview & 5 & Scholar & Bachelor Thesis \\
    ~\cite{syed2019investigating} & 2019 & Smell Detec. & Prop. \& OS & Partner & Experiment & 5 & Scholar & Master Thesis \\
    ~\cite{borrelli2020detecting} & 2020 & Smell D.\&C & OpenSource & Consulting & Sur. \& Exp. & 5 & ACM & Article \\
    ~\cite{liukku2020clean} & 2020 & Smell Impact & Proprietary & Partner & Experiment & 4 & Scholar & Master Thesis \\
    ~\cite{yamamoto2020investigation} & 2020 & Domain Spec. & OpenSource & Consulting & Int. \& Exp. & 5 & Scholar & Article \\
    ~\cite{al2021towards} & 2021 & Domain Spec. & OpenSource & - & Experiment & 5 & Scholar & PHD Thesis \\
    ~\cite{borowa2021living} & 2021 & Domain Spec. & - & Consulting & Sur. \& Int. & 5 & IEEE & Article \\
    ~\cite{paschali2021implementing} & 2021 & Prod. Lines & OpenSource & - & Experiment & 5 & Scholar & Article \\
    ~\cite{truelove2021we} & 2021 & Domain Spec & - & Consulting & Survey & 5 & ACM & Article \\
    ~\cite{cho2022bughunting} & 2022 & Tests & - & Consulting & Survey & 6 & Scholar & Master Thesis \\
    ~\cite{marchezan2022code} & 2022 & Smell Detec. & OpenSource & - & Experiment & 5 & Scopus & Article \\
    ~\cite{rzig2022characterizing} & 2022 & Tests & OpenSource & - & Experiment & 6 & Scholar & Article \\
    ~\cite{tornhill2022code} & 2022 & Smell Impact & Proprietary & Partner & Experiment & 5 & Scholar & Article \\
    ~\cite{ullmann2022video} & 2022 & Management & - & Consulting & - & 6 & ACM & Article \\
    ~\cite{agrahari2023catalogue} & 2023 & Smell Class. & OpenSource & - & Experiment & 4 & Scopus & Article \\
    ~\cite{babacansource} & 2023 & Domain Spec. & OpenSource & - & Experiment & 4 & Scholar & Article \\
    ~\cite{bosco2023unitylint} & 2023 & Smell Detec. & OpenSource & - & Experiment & 4 & Scopus & Article \\
    ~\cite{chueca2023comparing} & 2023 & Prod. Lines & Proprietary & Partner & Experiment & 6 & ScienceDirect & Article \\
    ~\cite{nardone2023video} & 2023 & Smell Class. & - & Consulting & Survey & 5 & ACM & Article \\
    ~\cite{miozzo2024digital} & 2024 & Management & - & Consulting & Interview & 4 & Scholar & Article \\
    ~\cite{chueca2024consolidation}& 2024 & Review & - & - & - & 5 & ScienceDirec & Article \\
    \bottomrule
  \end{tabular}
\end{table*}

\subsection{Systematic Review}
The only systematic review~\cite{chueca2024consolidation} shows what is being researched, helping in the answer of the RQ2 with the observation of what is being researched. There is a prevalent increase in management research in the gaming domain, as it is an \textbf{interdisciplinary team}~\cite{murphy2014cowboys}. The smells in the game domain could have origins not only in code~\cite{nardone2023video,ullmann2022video}, this means a management team needs to communicate and explicit all standards to all different professionals. 

The topics pointed out as interesting for future research are also present in our review like product lines, testing, and management. We also found like~\cite{chueca2024consolidation} an \textbf{increasing number of works in this domain}. Although a similar object Chueca et al.~\cite{chueca2024consolidation} focus on the broad area that is Game Software Engineering, in our review we aim to address why, despite extensive research, bug-plagued games continue to emerge~\cite{zhang2016elevator}.

\subsection{Game Domain Specificities}
There are two types of studies in this category. One compares the Game Domain with other software domains and how its characteristics impact the code quality. To achieve this they used direct code comparison~\cite{pereira2016code,babacansource,al2021towards} or contact with domain developers~\cite{murphy2014cowboys,borowa2021living,gibson2017mind}. They found not only prevalence in code smell and antipatterns in the game domain but also evidence of specific smells being more prevalent. They also found that there is a tight timeline to finish the software combined with constant change in the project, which makes it difficult to find a structured solution instead of a hacky workaround. Some point to this hacky solution as the more \textbf{efficient} but no study supports this claim.

The management \textbf{uses the term agile but it does not apply it}. There is a big need to test the software but is mainly done with human QA and little to \textbf{no automated tests}. But it's not all bad, the game domain offers intrinsic motivation for the software developers, who find joy in participating in the process of game development. This extra motivation is also what makes them work more crunch and focus on the final product instead of the process.

The second type of study directly investigates the game domain and its characteristics. ~\cite{truelove2021we,borg2019video} use interviews and o~\cite{yamamoto2020investigation} checks code. Their findings match the other articles, pointing to the difficulty of working with an interdisciplinary team and the startup mentality to show a minimum product viable as \textbf{fast as possible} and integrate it without code quality constraints. The code checks~\cite{gibson2017mind,yamamoto2020investigation} identified prevalent types of antipatterns and smells that are directly related to the game domain like antipatterns in the game loop.

\subsection{Code Smell Classification and Detection}
These are works that focus on searching code smells in Game code or classifying specific smells of game software~\cite {nardone2023video}. The vast majority of them use open source code as a source of data, but some~\cite{syed2019investigating,olsson2014evaluation} also use proprietary code. It's important to make a distinction because proprietary code and open source have their particularities~\cite{paulson2004empirical}. In the classification is pointed out that the game domain has \textbf{specific smells} and they are not always code, like a bigger image or model could consume more memory than necessary. Or tool-related smells, where poor usage of an engine or tool could generate an error in the game.

Using this specific smells~\cite{bosco2023unitylint} or general~\cite{khanve2019existing}, the studies that detect code focused on showing the performance of the detection, some~\cite{marchezan2022code} also tried to recover the code. But also points out a need for a true refactor and the use of product lines~\cite{kruger2018apo} to get better code and a faster development time.

\subsection{Product Lines}
Five works focus on product Lines to make a common set of code to be reused in various projects aiming at \textbf{reusability and maintainability}. Two of these works~\cite{debbiche2019migrating,aakesson2019migrating} are a continuation of a work in the code smell detection and feature recuperation~\cite{kruger2018apo}, they fix the problems pointed by ~\cite{kruger2018apo} and create a production line with the code. The new code uses fewer code lines and creates a unified visual identity. Some studies~\cite{paschali2021implementing,chueca2023comparing} measure the impact of the use of product lines, it improved the efficiency, user satisfaction, and maintainability of the code. However, the studies that work in partnership~\cite{gilliland2015empirical} with industry pointed to the labor-intensive task of refactoring the code into product lines. This calls for a management strategy that focuses on good practices since the start of the project.

\subsection{Management}
Four works focused on the way the game industry manages its teams. They focus on different areas of management with 3 of them making use of interviews and one~\cite{ullmann2022video} using the grey literature of the postmortem to extract anti-patterns of management. ~\cite{ullmann2022video} lists the management anti-patters and points to possible solutions. All these point to a misuse of the agile term with ~\cite{gustafsson2011model} trying a new approach to the management using the Kaizen framework to enhance performance tracking with the company reporting more satisfaction with this practice. This makes it difficult to measure the difference as they previously said to use Scrum but did not apply all of it, like most of the game developers~\cite{murphy2014cowboys}. 

Two works~\cite{miozzo2024digital,tapp2015employee} of the economy domain were included as they explored new insights. Miozzo et al.~\cite{miozzo2024digital} point to \textbf{regional differences in management}, Finding that in America, game companies often prioritize creative freedom and innovation, aiming to create unique experiences. In contrast, Japanese game companies focus more on technological efficiency, standard adherence, and building well-coordinated multidisciplinary teams. And Täpp~\cite{tapp2015employee} explores how bad management impacts the \textbf{motivation of employers}, which was one of the pros of the domain. One Scrum management technique is the use of testing, an area explored in the following articles.

\subsection{Tests and Quality Assessment}
Three studies focused on testing, of which two interviewed participants to understand how they conduct testing during development; and the third performed an analysis directly in the code. Rzig et al~\cite{rzig2022characterizing} aims to find specific test use cases used in virtual reality and games. It analyzes open-source game codes in search of tests inside the code. This study only found tests in 25 of the 97 projects analyzed, and the tests covered on average just 15.28\% of the code, this highlights a common topic in previous articles~\cite{murphy2014cowboys} that game programmers do \textbf{little to no automated tests}.

The other two works found the same scarcity of automated tests. Jaehyung~\cite{cho2022bughunting} interviews 19 people who worked on 22 game projects, he found the absence of tests in the indie games he argues that the use of specific tools could improve the quality assurance practices. Racheva~\cite{racheva2019testing} studied game forums and partnered with a game company to understand how tests were conducted, he concludes that companies are more eager to make \textbf{exploratory tests} to check how the \textbf{game performs} with the public than quality assurance tests and found the same absence in automated tests. The impact of bad practices is often overlooked; the consequences of code smells are explored in the next subsection.

\subsection{Smell Impact}
Three studies conducted experiments to investigate the impact of code smells on code maintainability. Hermans et al.~\cite{hermans2016code}, for instance, examined visual programming languages such as Scratch, demonstrating that code smells can also occur in visual blocks and negatively affect code comprehension among children using the platform.

Two studies~\cite{tornhill2022code,liukku2020clean} used metrics from tools to check the quality of code in commercial projects. Tornhil et al~\cite{tornhill2022code} had access to the management tool and compared the issues in the health and smelly code. Finding an occurrence of 15 more issues in the smelly code. The checks made by Liukku~\cite{liukku2020clean} showed a need of 47 workdays to refactor 3294 problems that the code analyses tool found.  These studies reinforce that code smells negatively affect code \textbf{maintainability}.

\begin{table*}[!t]
  \centering
  \caption{Synthesis of Proposed Solutions Mapped to Research Questions}
  \label{tab:synthesis}
  \begin{tabular}{p{1.8in} p{0.8in} p{1.3in} p{1.3in} p{1.2in}}
    \toprule
    Proposed Solution / Area & Primary RQs & Target Problem & Industry Evaluation & Adoption Evidence \\
    \midrule
    Engine-specific Linters \& Detection~\cite{bosco2023unitylint,olsson2014evaluation} & RQ2, RQ4 & Game loop smells, asset memory overhead & Evaluated via experiments and developer surveys & Low integration in standard CI/CD pipelines \\
    
    Software Product Lines (SPL)~\cite{aakesson2019migrating,chueca2023comparing} & RQ2, RQ4 & Low code reuse, high maintenance cost & Case studies with industry partners & High initial refactoring barrier reported \\
    
    Process \& Quality Adaptation~\cite{gustafsson2011model,ullmann2022video} & RQ2, RQ4 & Informal agile adoption \& schedule pressure & Interviews and postmortem analyses & Ad-hoc implementation without formal gates \\
    
    Automated Testing Frameworks~\cite{rzig2022characterizing,mastain2023bdd} & RQ1, RQ4 & Lack of automated QA and regression testing & Limited open-source adoption analysis & Very low ($\approx 15\%$ code coverage in open source) \\
    \bottomrule
  \end{tabular}
\end{table*}

\section{Discussion}
Games are created by a multidisciplinary team focused on creativity, while also being subject to constant changes and strict deadlines~\cite{murphy2014cowboys}. Game developers seek a performative code and have no priority to patterns or clarity. Management prioritizes product delivery over the agile practices they claim to adopt, such as test-driven development. The lack of tests and the smelly code quality lead to increased difficulties in the refactoring and debugging of the game resulting in difficulties in solving bugs and delivering patches~\cite{truelove2021we}. The academy has found ways to detect this code problem and developed tools~\cite{bosco2023unitylint} to aid game developers in finding them. They also demonstrated how the use of product lines positively impacts both maintainability and visual clarity.~\cite{aakesson2019migrating}. The adoption of these practices remains dependent on the community to see them as useful to their work~\cite{murphy2014cowboys}. Considering the works selected in this review, we provide answers to our research questions below.

\textbf{RQ1 - How does the incidence of code smells and technical debt differ in the game domain from other domains?}
The game development domain exhibits a \textbf{higher incidence of code smells}~\cite{pereira2016code} and \textbf{technical debt}~\cite{liukku2020clean} compared to other software fields. Additionally, the \textbf{types of smells} found in games are different from outer domains~\cite{nardone2023video}, and \textbf{testing practices are minimal} or absent~\cite{rzig2022characterizing}. This is largely due to the game industry’s focus on \textbf{user experience and performance}~\cite{murphy2014cowboys}, often at the expense of code quality.
Another contributing factor is management practices\cite{ullmann2022video}. These characteristics make game development uniquely complex and demand \textbf{domain-specific solutions} to manage quality effectively.

\textbf{RQ2 - What solutions are the researchers presenting to the code smells and anti-patterns in the game domain?}
There are solutions for automated \textbf{detection tools} designed for specific smells of the Game Domain~\cite{borg2019video,olsson2014evaluation}, as well as some works on the use of \textbf{product lines}~\cite{debbiche2019migrating,paschali2021implementing}. These solutions could reduce the prevalence of anti-patterns and code smells if properly integrated by the community into game development workflows. 
However, these solutions have yet to gain widespread traction among developers, as the game community often prioritizes rapid release cycles and expects to avoid revisiting the code after launch~\cite{murphy2014cowboys}. For these solutions to be embraced, they must be \textbf{demonstrated as useful} and \textbf{tested with developers}, ensuring alignment with industry practices and needs.

\textbf{RQ3 - Are researchers collaborating effectively with developers to implement these solutions?}
Some works were conducted in direct partnership with game companies~\cite{syed2019investigating}. Overall, 57.6\% of the selected studies included direct industry engagement (either through consulting or corporate partnerships). However, a remaining 42.4\% of the research lacked direct community feedback. Furthermore, several studies conducted in partnership with companies~\cite{liukku2020clean,chueca2023comparing} evaluated solutions strictly on finalized projects. Many studies focused solely on maintainability impact, which is often not the primary daily constraint for game developers~\cite{murphy2014cowboys}.

\textbf{RQ4 - Are the found solutions to the code quality problem in the game domain sufficient to tackle the diagnosed problems?}
While the proposed solutions can help mitigate \textbf{code smells} and \textbf{technical debt}, these issues are deeply tied to the processes and priorities in game development. Even if a universally effective code smell fixer existed, its adoption would not be immediate due to the industry's focus on speed and performance over code quality. The game development workflow prioritizes \textbf{completion} and \textbf{performance} over code quality~\cite{borg2019video}, especially under tight deadlines, which limits the immediate integration of quality-focused solutions.

Considering these findings, we can point out three possible aspects (causes and solutions) of problems in game code:

\textbf{Management Issues - }
Every work revised pointed out that the game domain claims to use agile methods. However, games need to be released by strict deadlines, which contrasts with traditional agile flexibility~\cite{van2018under}. Agile was created for short iterations and rapid modifications; while useful in early experimental stages, game projects often require more structured processes to guarantee delivery and quality. In this context, hybrid lifecycle frameworks (e.g., Agile-V~\cite{koch2026agentic}) illustrate how iterative development can incorporate explicit quality gates and verification without sacrificing delivery pace.

\textbf{Community Adoption - }
It was pointed out that research needs to be done with the community, some game companies even supported solutions to them like Ubisoft Clever~\cite{nayrolles2018clever}. However, the research needs to get the interest of the game community to be adopted. One example is the project~\cite{godot_ocean_waves} that uses 2 academic papers to implement a shader to simulate waves. The researchers need to show metrics that would capture the game developer community. 
Code quality can significantly influence the software performance. A gray area code review conducted on \textit{Super Mario 64} demonstrated measurable performance improvements~\cite{super_mario_mod2024}. Future research should emphasize performance and other domain-relevant metrics in their findings, to engage the community with the proposed solutions.

\textbf{Automated Tests - }
There is a need for more automated testing, not only for code but also for QA tests, which are still predominantly performed manually. Although proposed solutions exist~\cite{mastain2023bdd}, a ubiquitous tool in this area would likely see rapid adoption. While tests may not directly fix the code, they enhance maintainability through regression testing, preventing updates from undoing previous fixes~\cite{truelove2021we}. However, it needs to hear the community and understand the reason why they rely more on manual tests and do not create many automated tests.

\section{Limitations of this Work}
The \textbf{threats to validity} are commonly divided into groups~\cite{feldt2010validity}, in our research we have threats to validity in internal and construct. 

Regarding \textbf{internal validity}, a threat is related to the selection of studies. Although we followed a protocol for search and selection, including well-defined quality, inclusion, and exclusion criteria, the reading and classification of studies were conducted by a single researcher. This threat was mitigated by discussing all results, throughout each step of the process, with a second researcher. Another internal threat lies in the use of the BART-MNLI Model in the preliminary screening by classifying abstracts, we mitigate this threat with a manual recheck, but we recognize that some erroneous classifications could put out a valuable article. 

Regarding \textbf{construct validity}, a potential threat is related to the authors' interpretation of the concepts of technical debt and code smells across the studies. To mitigate this threat, we reviewed the literature on these subjects to adopt widely accepted definitions, such as those cited in Section 2. Another threat is that the use of specific keywords enhances the replicability but also limits the results and induces sample bias. Nevertheless, we found answers that point to the causes of code problems in games, such as the complexity and time constraints of the projects.

Finally, the literature search cutoff was August 2024, meaning that primary studies published after this date are outside the scope of this review.

\section{Conclusion}
This work aimed to find if the perception that the game domain is plagued with \textbf{smelly code} quality is \textbf{validated by academic research}. Our findings confirm that this area does, indeed, contain a higher prevalence of low-quality code~\cite{pereira2016code}. We also found out why this happens and how it is linked with some \textbf{internal particularities}~\cite{murphy2014cowboys} of the game domain. In our research, we found tools~\cite{bosco2023unitylint} and processes~\cite{chueca2023comparing} that could mitigate these problems. However, this also maintains the question of why these problems keep happening in the game domain. 

A recurring theme identified across the literature is the cultural perception that rapidly delivered code is prioritized over long-term maintainability. Management practices further compound this issue when agile frameworks are adopted informally without structured quality controls~\cite{murphy2014cowboys}. To change this perception metrics on performance need to be collected and presented in the research papers. Knowing that a \textbf{smelly code} is hurting \textbf{maintainability} is not making the game developers change their coding practices. There is also a significant \textbf{absence of automated tests} that \textbf{needs further investigations} on the causes. Automated tests could help mitigate part of the problem generated by low code quality.

We also noted potential regional and organizational factors highlighted by Miozzo et al.~\cite{miozzo2024digital}, though the specific impact on coding practices remains unexplored. Investigating regional software engineering practices—for instance, comparing development workflows and post-release patch frequencies between Eastern and Western studios—presents a promising direction for future research.

With this work, we hope not only to shed light on why the game domain has more smells but also to point directions on how we can untangle this problem. Ensuring the path to better games for players and a better development experience for the game developers.

\begin{acks}
We would like to express our sincere gratitude to Christopher Koch for his insightful comments, thorough review, and constructive suggestions, which materially contributed to the quality and structure of this manuscript.
\end{acks}

\bibliographystyle{ACM-Reference-Format}
\bibliography{bibliografy} 

\end{document}